\documentclass{elsart5p}

\usepackage{graphics}
\usepackage{graphicx}
\usepackage{amssymb}
\usepackage{epstopdf}
  \def \In {$^{115}$In }
    \def \Ce {CeCoIn$_5 $ }

\begin{document}

\begin{frontmatter}



\title{Phase Diagram of CeCoIn$_5$ in the Vicinity of $H_{c2}$ as Determined by NMR}
%

\author[AA]{V. F. Mitrovi{\'c}\corauthref{VFM}},
\ead{vemi@physics.brown.edu}
\author[AA]{G. Koutroulakis},
\author[AA]{M. -A. Vachon},
\author[BB]{M. Horvati{\'c}},
\author[BB]{C. Berthier},
\author[CC]{G. Lapertot},
\author[CC]{J. Flouquet},

\address[AA]{Department of Physics, Brown University, Providence, RI 02912, U.S.A. }
\address[BB]{Grenoble High Magnetic Field Laboratory, CNRS, B.P. 166, 38042 Grenoble
Cedex 9, France}
\address[CC]{D{\'e}partement de Recherche Fondamentale sur la Mati{\`e}re Condens{\'e}e, SPSMS, CEA Grenoble, 38054 Grenoble Cedex 9, France}

\corauth[VFM]{Corresponding author. Tel: (1 401) 863-2587 fax: (1 401)
863-3359}

\begin{abstract}
We report  \In nuclear magnetic resonance (NMR)  measurements in   the heavy-fermion superconductor \Ce  as a  function of    temperature in different magnetic fields applied parallel to the $(\hat a, \hat b)$ plane.  The measurements  probe a part of the phase diagram in the vicinity of the superconducting critical field $H_{c2}$ where a possible inhomogeneous superconducting state,  Fulde-Ferrel-Larkin-Ovchinnikov (FFLO),  is stabilized. We have identified clear NMR signatures of two phase transitions occurring in this part of the phase diagram.  The first order phase transitions are characterized by the sizable discontinuity of the shift. 
We find that a continuous second order phase transition from the superconducting to the FFLO state 
  occurs at temperature below which the shift becomes temperature independent. We have compiled the first phase diagram  of \Ce in the vicinity of $H_{c2}$ from   NMR data and found that it is in 
 agreement with the one determined   by thermodynamic measurements. 
\end{abstract}


\end{frontmatter}


\section{Introduction}

Very soon after the development of the BCS theory there have been many efforts to go beyond conventional superconductivity investigating other possible superconducting phases. One of the most notable examples is the prediction of Fulde, Ferrell,  Larkin, and Ovchinikov (FFLO) \cite{FFLOdis} of an unconventional superconducting state realized in the vicinity of the critical field $H_{c_2}$. In general, superconductivity is suppressed in the presence of an applied magnetic field $(H_0)$ through two distinct physical mechanisms: one, the coupling of the magnetic field to the electronic orbital angular momentum which leads to the formation of a regular vortex array (Abrikosov vortex state); two, the Cooper pairs breaking caused by the Zeeman interaction between the field and the electron spins (Pauli paramagnetism). FFLO argued that when the Pauli pair breaking dominates over the orbital effects, the normal state is unstable with respect to the formation of a new type of electron pairs with non-zero center-of-mass momentum. Specifically, a pairing state $\left(\textbf{k}_{\uparrow}, -\textbf{k}+\textbf{q}_\downarrow \right)$ with $q\sim2\mu_BH_0/\hbar v_F$ 
is stabilized reducing the Pauli paramagnetic effect. The appearance of the non-zero momentum $q$ breaks the spatial symmetry and, consequently, an inhomogeneous superconducting phase with   order parameter that oscillates in real space  emerges. It is striking that despite the early theoretical prediction of this exotic superconducting phase its commonly accepted experimental realization is yet to be observed. 

Extensive efforts have been made to detect the FFLO state in materials so  that the rich physics of the state can be explored.     The material properties that are favorable to formation of the FFLO state include  high purity  ($l \gg \xi$), large spin susceptibility favoring   spin coupling to the applied field, and layered structure that inhibits electronic orbital motion for magnetic fields parallel to the conducting planes.   
Single crystals of \Ce are identified as good candidate  for   formation of the FFLO. Indeed, many experiments \cite{Bianchi03, Martin05, mitrovic06, Young06}  have identified a possible phase transition to a FFLO state. The phase transition is discerned within the Abrikosov superconducting (SC) state (henceforth referred to as  SC state)
    in the vicinity of the $H_{c2}$. 

 Most experiments have focused on the detection of a phase transition between the  SC and FFLO states and/or between the FFLO  and normal states.  
However, direct evidence of the order parameter oscillation in the real space is still lacking and the  nature of the   new low temperature SC state  thus       remains to be elucidated. NMR appears to be the technique of choice for probing the microscopic nature of the  state.    
 However, clear signature of the transition to the FFLO state from SC state across continuous second order transition is still lacking.   To address this issue we have performed extended  measurements of  the temperature dependence of the NMR shift   throughout the high field part of the  phase diagram.

Here we report \In NMR measurements  as a function of  temperature in various applied magnetic fields 
in the vicinity of $H_{c2}$  on a single crystal sample of  CeCoIn$_5$. 
 The shift and the lineshape measurements probe the local spin susceptibility  ($\chi$) and 
  a local magnetic field map,  respectively.  By examining the temperature ($T$) dependence of the 
  shift, we identify clear NMR signatures  of two phase transitions at low temperatures and high magnetic fields. These are: the sizable discontinuity of the shift at the first order phase transition and 
  onset of the  temperature independence of the shift at the 
  continuous second order phase transition from the superconducting to the FFLO state.  
   Furthermore, we show that the phase transitions are accompanied by concomitant line broadening below the transition temperatures. We have compiled the phase diagram   in the vicinity of $H_{c2}$ and found it in agreement with  the one determined   by thermodynamic measurements \cite{Bianchi03}.

 \section{Experimental Technique and the Sample}
   
   We have used high quality single crystals of \Ce  grown by a flux method   \mbox{\cite{CedaDic}}. 
 The  In $(I=9/2)$ quadrupolar-split NMR satellite lines were used to infer that $H_0$ was aligned to better than 1$^\circ$  with respect to the sample's   $(\hat a \hat b)$ plane.  Here, the spectra  of  the axially symmetric $^{115}$In(1) site  are reported.  They   were recorded using a custom built NMR spectrometer and obtained, at each given value of $H_0$, from the sum of spin-echo Fourier transforms recorded at  each 10 (or 20) KHz interval.     
 The shift 
was determined by the diagonalization of the nuclear spin 
Hamiltonian including the quadrupolar effects. Pure  magnetic contribution of the shift is obtained after subtracting   independent orbital contribution of 0.13\% to the 
shift.  
  
 The low temperature environment  was provided by a $^3$He/$^4$He dilution refrigerator. 
   The RF coil was mounted into the mixing chamber of the refrigerator,  
   ensuring  good thermal contact.  We used both  the `top-tuning' and   fixed  `bottom-tuning' scheme. The `top-tuning' scheme, in which the  variable tuning delay line and matching capacitor were mounted outside the NMR probe, enabled us 
  to span over a wide field/frequency range, while the   `bottom-tuning', in which a fixed capacitor was mounted in the mixing  chamber near the coil, was used to assure improved signal-to-noise ratio.  The sample was both  zero-field and field-cooled. No notable influence  of the sample cooling history on the NMR shift  was detected. 
 In order  to avoid  heating of    the sample by  RF pulses we used an 
  RF excitation power much weaker than  usual and  repetition times of the order of 
 a second to several seconds, depending on temperature.

\section{Magnetic Shift}

\begin{figure}[b]
   \begin{minipage}{1.02\hsize}
\centerline{\includegraphics[scale=0.49]{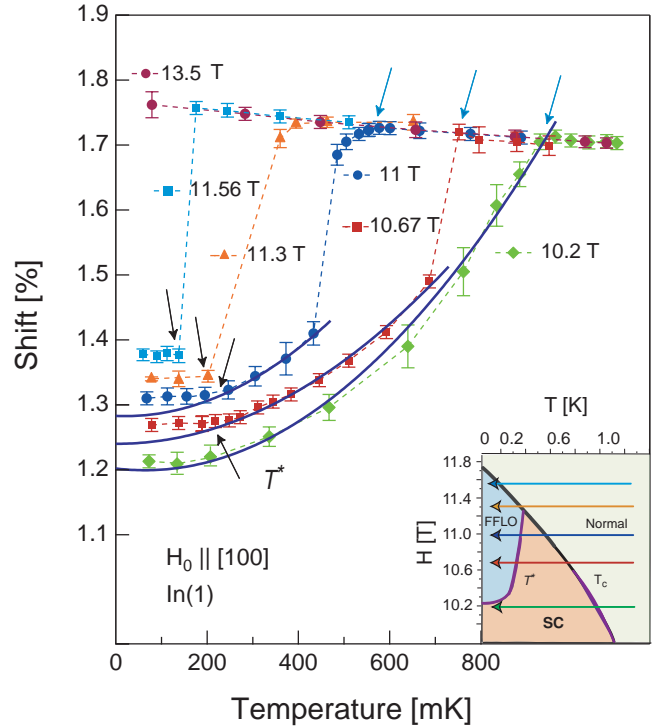}} 
\begin{minipage}{0.98\hsize}
\caption[]{\label{Fig1}\small The magnetic shift of $^{115}$In  as a function of $T$ at different  $H ||[100]$, as indicated. The dashed lines are   guides to  the eye.  Solid lines denote the $T^2$ dependence. 
Typical error bars are shown.  
The blue arrows  indicate $T_c$,  the transition temperature  to the SC   state. The black arrows indicate approximate location of $T^*$,  the transition temperature  to the supposed FFLO state. Inset:     Sketch of the $H - T$ phase diagram  of \Ce for $H_0 \bot \hat c$. Black and purple lines indicate the  $1^{\rm st}$  and  $2^{\rm nd}$  order SC phase transitions, respectively. Arrows indicate the fields in which spectra in the figure are obtained. 
}
 \end{minipage}
 \end{minipage}
\end{figure}

The temperature dependence of the shift at different applied fields is shown in \mbox{Fig. \ref{Fig1}}. 
 The region of the phase diagram that was scanned in these measurements is indicated by arrows in  the inset to the figure.    In fields ranging from  10.2  T to 12 T, the $T$ dependence of  shift clearly reveals two   phase transitions,  at $T_c$ and $T^*$,    as indicated in   \mbox{Fig. \ref{Fig1}}  by the blue and black arrows, respectively.  
   The shift is discontinuous across the first-order  and varies smoothly across the second-order phase transitions \cite{Bianchi03}.     Even though, the NMR measurements cannot  {\it a priori} unambiguously determine the order of a phase transition, 
   in this case the NMR shift  correctly  reveals   the order. 
     As was previously discussed  \cite{mitrovic06},
  the     phase transition at  $T_c$   can be clearly associated with the transition to the uniform SC   state. 
 This conclusion comes from the observation of   the strong  loss of signal intensity   due to RF shielding by  SC currents.

 Furthermore, three distinct temperature dependence of the shift are evident. Thus, the distinct  $T$ variation of the shift can be interpreted as a signature of separate states stabilized in this part of the phase diagram.  
   The normal state shift exhibits linear $T$ dependence and increases with decreasing temperature. 
    In the SC state the  temperature  dependence of the shift,  is expected to be quadratic, when 
    $\mu_B H_0 > k_BT$ \cite{AntonLT, KunYang}.  Indeed,  the quadratic $T$ dependence is observed at 10.2 T, 10.67 T, and 11 T fields, as indicated by solid lines in  \mbox{Fig. \ref{Fig1}}. 
   However,  at 10.67 T and 11 T at low temperatures 
  there is an obvious discrepancy, well outside the error bars, between data   and the expected 
$T$ dependence in the   SC state. We point out that the difference is particularly significant on the reduced $T$ scale as was shown in {Ref. \cite{mitrovic06}}
 Moreover,  shift is temperature independent in this low temperature part of the phase diagram. We interpret this $T$ independence of $\chi$ as a signature of the FFLO state.

Two observations support our inference that the signature of the FFLO state is the $T$ independent shift. First, in $H_0 > 11$ T   the transition from the normal to the FFLO state  is expected to be of  the  first order  as per specific heat measurements \cite{Bianchi03}. At $H_0 = 11.3$ T and 11.56 T, shift is temperature independent below  the $T$ at which it  discontinuously  decreases from its  normal state value. For  $H_0 \le 11$ T  temperature independence of the shift onsets concomitantly with the departure    from its expected $T^2$ behavior. 
    
   The  $T$ independence of the shift supports the FFLO scenario for this phase as opposed to the magnetic one \cite{Young06}. Obviously, the low temperature  shift exhibits a strong field dependence. Therefore, in the magnetic scenario   it would be hard to conciliate the fact that moments do not respond to temperature perturbation   while the system is readily affected by the changes of $H_0$.     

\begin{figure}[b]
   \begin{minipage}{0.98\hsize}
\centerline{\includegraphics[scale=0.49]{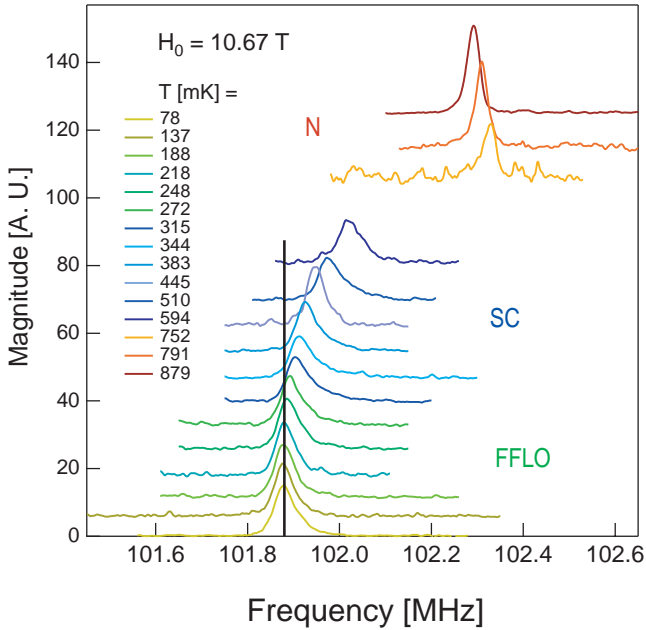}} 
\begin{minipage}{0.98\hsize}
\caption[]{\label{Fig2}\small $^{15}$In(1) central transition spectra as a function of temperature at 10.67 T magnetic field applied in the [100] direction. Solid line is  guide to the eye that indicates the frequency position corresponding to temperature independent shift in the FFLO state.}
 \end{minipage}
 \end{minipage}
\end{figure}

\section{NMR Lineshape in the Possible FFLO State}

To further address the issue of the NMR signature of the FFLO
phase, we examine the NMR lineshape at 10.67 T whose temperature dependence is plotted in \mbox{Fig. \ref{Fig2}}.
 The solid line in the figure indicates the frequency position corresponding to temperature independent shift in the FFLO state. Peak position of all green shaded spectra coincides with the line. 
  Above $T \approx 250$ mK, peak of the spectra shifts to higher frequencies with increasing temperature. 
 \begin{figure}[t]
    \begin{minipage}{1.05\hsize}
\centerline{\includegraphics[scale=0.43]{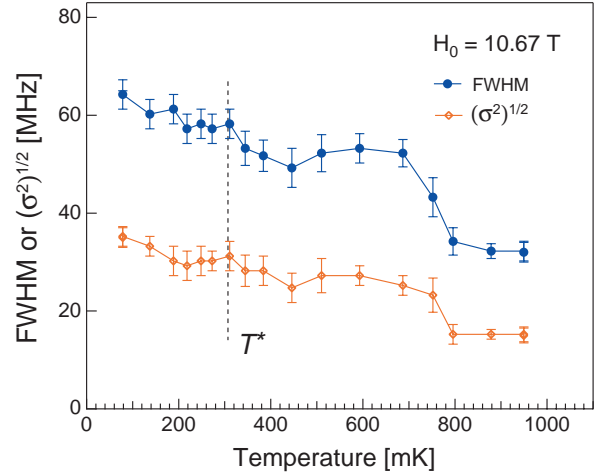}} 
\begin{minipage}{0.98\hsize}
\caption[]{\label{Fig3}\small 
The full-width-half-maximum (FWHM) ($\bullet$) and square root of the second moment ($\sqrt{\sigma^2}$) ($\diamond$) of the measured \In spectra   as a function of temperature at 10.67 T magnetic field, $H_0 || [100]$.  
The solid lines are guides  to  the eye. }
 \end{minipage}
\end{minipage}
\end{figure}

Let us now consider the $T$ dependence of
the lineshape. The full width at the half maximum (FWHM) and square root of the second moment ($\sqrt{\sigma^2}$) of the    spectra at 10.67 T are plotted 
in  \mbox{Fig. \ref{Fig3}}  as a function of $T$. Since
$T^{*}$ value determined from the $T$ dependence of the shift is
less than $T_{c}/2$,  one expects that the broadening due to the
vortex lattice should become temperature independent below
$T^{*}$. On the contrary, the second moment continuously
increases, due to the entry in a new phase.
 Extra broadening in the FFLO phase is indeed expected due to the spatial modulation of magnetization in the phase.  
 The absence of discontinuity of the second moment
$\sigma^{2}$ at $T^{*}$ is in
agreement with a second order transition.

\section{High field Low Temperature Phase Diagram as Determined by NMR}

\begin{figure}[t]
    \begin{minipage}{0.97\hsize}
\centerline{\includegraphics[scale=0.57]{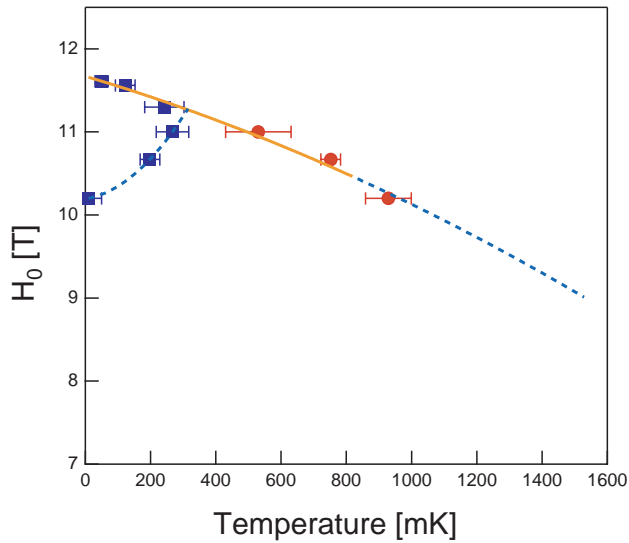}} 
\begin{minipage}{0.98\hsize}
\caption[]{\label{Fig4}\small 
Phase diagram  of \Ce for $H_0 \bot \hat c$ as compiled from the NMR measurement presented in this paper. Red circles denote $T_c$, transition temperature from the normal to the SC state. Blue squares 
mark $T^*$, transition temperature to a possible FFLO state. Solid and dashed line indicates   the first  and the second order phase transition, respectively.}
 \end{minipage}
\end{minipage}
\end{figure}

 A phase diagram as determined by our NMR measurements is plotted in  \mbox{Fig. \ref{Fig4}}. 
It is in excellent agreement with the one determined by thermodynamic measurements   of  A. Bianchi {\it et al.} \cite{Bianchi03}.

\section{Summary}
In conclusion, we   measured the NMR  spectra  of In(1)  in a single crystal sample of  \Ce   as a function of    temperature in different magnetic fields, applied parallel to SC  planes,  in the vicinity of the upper critical field.   
Clear  NMR signatures of two phase transitions at low temperatures and high magnetic fields are  identified. The first order phase transitions are characterized by the sizable discontinuity of the shift. 
We find that a continuous second order phase transition from the superconducting to the FFLO state 
  occurs at temperature below which the shift becomes temperature independent, {i.e.} the temperature dependence of the shifts departs from its characteristic $T^2$ behavior.  Furthermore, we show that the phase transitions are accompanied by concomitant line broadening below the transition temperatures. We have compiled the first phase diagram of \Ce in the vicinity of $H_{c2}$ from NMR measurements. 
The phase diagram is in very good agreement with the one determined   by thermodynamic measurements.

\section{Acknowledgement}
 We thank S. Kr{\"a}mer for his contribution. 
This research is supported in part by the funds from  
 the National Science Foundation    (DMR-0547938), 
Brown University,  and the Grenoble High Magnetic Field Laboratory, under European Community contract RITA-CT-2003-505474.


\begin{thebibliography}{99}
\bibitem{FFLOdis} 
 P. Fulde and R.A. Ferrell, Phys. Rev. {\bf 135} (1964) A550; A.I. Larkin and Y.N. Ovchinnikov, Zh. Eksp. Teor. Fiz. {\bf 47} (1964) 1136.  
 \bibitem{Bianchi03}
 A. Bianchi {\it et al.}, Phys. Rev. Lett. {\bf 91}  (2003) 187004.
\bibitem{Martin05}
 C. Martin  {\it et al.}, Phys. Rev. B {\bf 71} (2005) 020503 .
\bibitem{mitrovic06}
V. F. Mitrovi{\' c} {\it et al.}, Phys. Rev. Lett. {\bf 97}  (2006) 117002.
\bibitem{Young06}
B.-L. Young {\it et al.}, Phys. Rev. Lett. {\bf 98}  (2006) 036402.
  \bibitem{CedaDic}
{\v C}. Petrovi{\' c}  {\it et al.},  J. Phys. Condens. Matt.  {\bf 13}  (2001) L337.
\bibitem{AntonLT}
A. B. Vorontsov and M. J. Graf, {\it cond-mat} 0507479.
\bibitem{KunYang} 
Kun Yang and S. L. Sondhi,  Phys. Rev. B  {\bf 57}, 8566 (1998).

 
 
\end{thebibliography}
\end{document}